\documentclass[sigconf,authorversion]{acmart}

\usepackage{booktabs} 
\usepackage{enumitem}
\usepackage[skip=0pt]{caption}
\usepackage{url}

\setcopyright{rightsretained}

\copyrightyear{2017}
\acmYear{2017}
\setcopyright{acmlicensed}
\acmConference{SIGIR '17}{August 07-11, 2017}{Shinjuku, Tokyo,
Japan}
\acmPrice{15.00}
\acmDOI{10.1145/3077136.3082062}
\acmISBN{978-1-4503-5022-8/17/08}

\usepackage{flushend}
\usepackage{multirow}

\setlist[description]{leftmargin=0cm,labelindent=0cm}

\newcommand{\TomKenter}{TK}
\newcommand{\AlexeyBorisov}{AB}
\newcommand{\MostafaDehghani}{MD}
\newcommand{\BhaskarMitra}{BM}
\newcommand{\ChristopheVanGysel}{CVG}
\newcommand{\MaartendeRijke}{MdR}

\begin{document}

\title{Neural Networks for Information Retrieval}

\author{Tom Kenter}
\authornote{Corresponding author.}
\affiliation{%
  \institution{University of Amsterdam}
  \city{Amsterdam, The Netherlands}
}
\email{tom.kenter@uva.nl}

\author{Alexey Borisov}
\affiliation{%
  \institution{Yandex}
  \city{Moscow, Russia}
}
\email{alborisov@yandex-team.ru}

\author{Christophe Van Gysel}
\affiliation{%
  \institution{University of Amsterdam}
  \city{Amsterdam, The Netherlands}
}
\email{cvangysel@uva.nl}

\author{Mostafa Dehghani}
\affiliation{%
  \institution{University of Amsterdam}
  \city{Amsterdam, The Netherlands}
}
\email{dehghani@uva.nl}

\author{Maarten de Rijke}
\affiliation{%
  \institution{University of Amsterdam}
  \city{Amsterdam, The Netherlands}
}
\email{derijke@uva.nl}

\author{Bhaskar Mitra}
\affiliation{%
  \institution{Microsoft, University College London}
  \city{Cambridge, UK}
}
\email{bmitra@microsoft.com}

\renewcommand{\shortauthors}{T. Kenter et al.}


\maketitle

\section*{Extended abstract}

Machine learning plays a role in many aspects of modern IR systems, and deep learning is applied in all of them.
The fast pace of modern-day research has given rise to many different approaches for many different IR problems.
The amount of information available can be overwhelming both for junior students and for experienced researchers looking for new research topics and directions.
Additionally, it is interesting to see what key insights into IR problems the new technologies are able to give us.
The aim of this full-day tutorial is to give a clear overview of current tried-and-trusted neural methods in IR and how they benefit IR research.
It covers key architectures, as well as the most promising future directions.


\section{Motivation}

Prompted by the advances of deep learning in computer vision research, neural networks have resurfaced as a popular machine learning paradigm in many other directions of research as well, including information retrieval.
Recent years have seen neural networks being applied to all key parts of the typical modern IR pipeline, such core ranking algorithms \cite{szummer2011semi,huang2013learning,Mitra:2016:www}, click models \citep{Borisov:2016:click,Borisov:2016:time}, knowledge graphs \cite{bordes2011learning,lin2015learning}, text similarity \cite{kenter2016siamesecbow,severyn2015learning}, entity retrieval \citep{VanGysel:2016:www,VanGysel:2016:cikm}, language modeling \cite{bengio2003neural}, question answering \cite{weston2015towards,hewlett2016wikireading}, and dialogue systems \cite{li2016deep,vinyals2015neural}.

A key advantage that sets neural networks apart from many learning strategies employed earlier, is their ability to work from raw input data.
E.g., when given enough training data, well-designed networks can become feature extractors themselves, e.g., incorporating basic input characteristics such as term frequency (tf) and term saliency (idf)---that used to be pre-calculated offline---in their initial layers.
Where designing features used to be a crucial aspect and contribution of newly proposed IR approaches, the focus has shifted to designing network architectures instead.
As a consequence, many different architectures and paradigms have been proposed, such as auto-encoders, recursive networks, recurrent networks, convolutional networks, various embedding methods, deep reinforcement and deep q-learning, and, more recently, generative adversarial networks, of which most have been applied in IR settings.
The aim of the \textit{neural networks for IR (NN4IR)} tutorial is to provide a clear overview of the main network architectures currently applied in IR and to show explicitly how they relate to previous work.
The tutorial covers methods applied in industry and academia, with in-depth insights into the underlying theory, core IR tasks, applicability, key assets and handicaps, scalability concerns and practical tips and tricks.

We expect the tutorial to be useful both for academic and industrial researchers and practitioners who either want to develop new neural models, use them in their own research in other areas or apply the models described here to improve actual IR systems.


\section{Objectives}

The material in the tutorial covers a broad range of IR applications.
It is structured as follows: 


\paragraph{\textbf{Preliminaries} (60 minutes)}

The recent surge of interest in deep learning has given rise to a myriad of architectures.
Different though the inner structures of neural networks can be, there are many concepts common to all of them.  
This first session covers the preliminaries; we briefly recapitulate the basic concepts involved in neural systems, such as back propagation \cite{rumelhart1988learning}, distributed representations/embeddings \cite{mikolov2013efficient}, convolutional layers \cite{krizhevsky2012imagenet}, recurrent networks \cite{mikolov2010recurrent}, sequence-to-sequence models \cite{sutskever2014sequence}, dropout \cite{srivastava2014dropout}, loss functions, optimization schemes like Adam \cite{kingma2014adam}.

\paragraph{\textbf{Semantic matching I: supervised learning} (60 minutes)}

The problem of matching items based on their textual descriptions arises in many IR systems.
The traditional approach involves counting query term occurrences in the description text (e.g., BM25 \cite{robertson1995okapi}).
However, to bridge the lexical gap caused by vocabulary-related and linguistic differences many latent semantic models have been proposed \cite{deerwester1990indexing, hofmann1999probabilistic, blei2003latent, wei2006lda}, and more recently neural embedding methods \cite{mikolov2013efficient}.
In this session we will focus on semantic matching settings where a supervised signal is available.
The signal can be explicit, such as a label for learning task-specific latent representations \cite{lu2013deep,huang2013learning, shen2014latent, hu2014convolutional, severyn2015learning,kenter2015short}, or relevance labels and, more implicitly, clicks for neural IR methods \cite{mitra2016desm, diaz2016query, grbovic2015context, kusner2015word,Mitra:2016:www}.

\paragraph{\textbf{Semantic matching II: Semi- and unsupervised learning} (60 minutes)}

How to learn semantics in the absence of relevance labels or user interaction signals? Depending on the available resources, one can choose for semi- or unsupervised matching models. 
 
Unsupervised semantic matching methods can be categorized into two groups. First, \emph{using pre-trained word embeddings} like combining traditional retrieval models with an embedding-based translation model~\citep{Zuccon2015nntm, Ganguly2015generalizedlm}, using pre-trained embeddings for query expansion to improve retrieval~\citep{Zamani2016queryexpansion}, and representing documents as Bag-of-Word-Embeddings (BoWE)~\citep{Guo2016wordtransport,kenter2015short}.
Second, \emph{learning representations from scratch} like learning representations of words and documents \citep{le2014distributed,kenter2016siamesecbow} and employing them in retrieval task~\citep{Ai2016doc2veclm,Ai2016doc2vecanalysis}, and learning representations in an end-to-end neural model for learning a specific task like entity ranking for expert finding~\citep{VanGysel:2016:www} or product search~\citep{VanGysel:2016:cikm}.

In semi-supervised learning, on the other hand, queries (without relevance labels), or prior knowledge about document similarity can be used to induce pseudo-relevance labels.
Furthermore, it is possible to use heuristic methods to generate weak supervision signals and to go beyond them by employing proper learning objectives and network designs~\citep{Dehghani:2017}.


\paragraph{\textbf{Learning to rank} (45 minutes)}

Capturing the notion of relevance for ranking needs to account for different aspects of the query, the document, and their relationship.
Neural methods for ranking can use both manually crafted features from query and document and combine them with regards to a ranking objective, or learn latent representations for them in situ. 

Irrespective of how the query and the documents are featurized, a neural learning to rank model can be designed for different scenarios, each having its own appropriate loss function.
An example is the point-wise versus pair-wise paradigm, each of which has a different objective that calibrates either scores or the relative ranking of documents, given a query.
Neural learning to rank models can also be designed to be provided with different levels of supervision during training---unsupervised \cite{VanGysel:2016:www, VanGysel:2016:cikm, salakhutdinov2009semantic}, semi/weakly-supervised \cite{Dehghani:2017, szummer2011semi}, or fully-supervised using labeled \cite{Mitra:2016:www} or click data \cite{huang2013learning}.

\paragraph{\textbf{Modeling user behavior} (45 minutes)}

Modeling user browsing behavior plays an important role in the development of modern IR systems. Accurately interpreting user clicks is difficult due to various types of bias. For example, users tend to click more on results ranked on top positions (position bias) and visually salient results (attention bias). The traditional way to account for these biases is to design a Probabilistic Graphical Model (PGM) that explains relationships between click/skip events (observed variables) and examination (unobserved variables). Over the last decade many PGM-based click models have been proposed (see \cite{chuklin2015click} for an overview). However, these click models can model only those patterns that are explicitly encoded in their PGMs.
Recently, it was shown that recurrent neural networks can learn to account for biases in user clicks directly from the click-through data, i.e., without the need for a predefined set of rules as is customary for PGM-based click models \cite{Borisov:2016:click}.
Additionally, there are similar biases in click dwell times, which the neural approach can account for too.


\paragraph{\textbf{Generating responses} (45 minutes)}

Recent inventions such as smart home devices, voice search and virtual assistants provide new ways of accessing information. They require a different response format than the classic ten blue links.
Targeting this newly emerging demand, some models have been proposed to respond by generating natural language replies on the fly, rather than by (re)ranking a fixed set of items or extracting passages from existing pages.

Examples are conversational and dialog systems \citep{li2016deep,vinyals2015neural, bordes:2016:learning} or machine reading and question answering tasks where the model either infers the answer from unstructured data, like textual documents that do not necessarily feature the answer literally~\citep{Hermann:2015:nips, weston2015towards,hewlett2016wikireading, serban2016generating}, or generates natural language given structured data, like data from knowledge graphs or from external memories~\citep{Ahn:2017,Lebret:2016, mei:2015, miller2016key,Graves:2014:neural}.

\paragraph{\textbf{Outlook} (30 minutes)}

In this session, open research questions and future directions are discussed.
One of the big challenges for IR at the moment is how to process full document text using neural networks.
On a higher level, it is probably desirable to learn all components of a full IR system in an end-to-end fashion.

Another challenge is maintaining long term (multiple day) search sessions or conversations.
Which naturally leads to an additional open problem: how to evaluate (neural) conversational systems.

Finally, we cover recent advances, like Generative Adversarial Networks \cite{goodfellow2014generative}.

\medskip\noindent%
Summing up, the \textbf{objectives of the NN4IR tutorial} are as follows:
\begin{itemize}[leftmargin=0.13in]
\item Give an extensive overview of neural network architectures currently employed in IR, both in academia and industry.
\item Provide theoretical background, thereby equipping participants with the necessary means to form intuitions about various neural methods and their applicability.
\item Identify the IR lessons learned by employing neural methods.
\item Give practical tips and tricks, regarding network design, optimization, hyperparameter values, based on industry best practice. 
\item Discuss promising future research directions.
\end{itemize}

\noindent%
The \textbf{target audience} consists of researchers and developers in information retrieval who are interested in gaining an in-depth understanding of neural models across a wide range of IR problems.
The tutorial will be useful as an overview for anyone new to the deep learning field as well as for practitioners seeking concrete recipes.
The tutorial aims to provide a map of the increasingly rich landscape of neural models in IR.

By the end of the tutorial, attendees will be familiar with the main architectures of neural networks as applied in IR and they will have informed intuitions of their key properties and of the insights they bring into core IR problems.
We aim to provide an overview of the main directions currently employed, together with a clear understanding of the underlying theory and insights, illustrated with examples.


\section{Format and detailed schedule}
Table~\ref{table:schedule} gives an overview of the time schedule of the tutorial.
Below we provide the details for each session. 

\begin{table}[t]
  \centering
  \caption{Time schedule for NN4IR tutorial}
  \label{table:schedule}
  \begin{tabular}{l l}
    \toprule
    Preliminaries          & 60 minutes \\
    Semantic Matching I    & 60 minutes \\
    Semantic Matching II   & 60 minutes \\
    \midrule
    \multicolumn{2}{c}{\textit{lunch break}} \\
    \midrule
    Learning to Rank       & 45 minutes \\
    Modeling User Behavior & 45 minutes \\
    Generating Responses   & 45 minutes \\
    Outlook                & 30 minutes \\
    Wrap up                & 15 minutes \\
    \bottomrule
  \end{tabular}
  \vspace*{-\baselineskip}
\end{table}

We bring a team team of six lecturers, all with their specific areas of specialization. Each session will have two expert lecturers who will together present the session.
The initials below refer to the lecturers for this tutorial.


\paragraph{Preliminaries --- 60 minutes (\TomKenter, \MaartendeRijke)}
\begin{enumerate}[itemindent=25pt,leftmargin=0.15in]
\item[\textbf{10 mins}] Back propagation --
                Given a standard feedforward network, we show the math and the intuition of back propagation. Also we will briefly touch on dropout.
\item[\textbf{5 mins}]  Distributed representations --
                We show what a distributed representation is, and how distributed representations can be used. 
\item[\textbf{10 mins}] Recurrent neural networks --
                 We cover the basics, based on a language modeling scenario, including LSTMs \cite{hochreiter1997long} and GRUs \cite{cho2014learning}.
\item[\textbf{10 mins}] Embedding methods  --
                 We detail how word2vec works and how it can be applied to different settings. 
\item[\textbf{10 mins}] Sequence-to-sequence models --
                 Basic architecture of seq2seq models \cite{sutskever2014sequence}, including the attention mechanism \cite{bahdanau2014neural}.
\item[\textbf{10 mins}] Convolutional networks --
                 CNNs are primarily employed in computer vision, but can be beneficial in text classification tasks too.
\item[\textbf{5 mins}] Optimisation schemes --
                 Standard back propagation with a fixed learning rate is typically replaced by more sophisticated schemes that handle learning rate annealing.
\end{enumerate}


\paragraph{Semantic matching I: supervised --- 60 minutes (\AlexeyBorisov, \BhaskarMitra)}

\begin{enumerate}[itemindent=25pt,leftmargin=0.15in]
\item[\textbf{10 mins}] Short text similarity --
                 Given two short texts, e.g., queries or sentences, how can we predict if they are semantically similar?
\item[\textbf{15 mins}] Word embeddings for matching --
                 Learning embeddings from click data.
\item[\textbf{20 mins}] Deep neural architectures for matching --
                 Deep Structured Semantic Model (DSSM) \cite{huang2013learning}.\item[\textbf{15 mins}] Learning to match using local representations --
                 Use both local and global representations for query-document matching.
\end{enumerate}


\paragraph{Semantic matching II: Semi- and unsupervised semantic matching --- 60 minutes (\MostafaDehghani, \ChristopheVanGysel)}

\begin{enumerate}[itemindent=25pt,leftmargin=0.15in]
\item[\textbf{10 mins}] Semi-supervised semantic matching --
We cover how to model pseudo-labeling using prior knowledge like document similarity, or by employing heuristic methods as weak supervision signals.

\item[\textbf{25 mins}] Unsupervised semantic matching using pre-trained word embeddings --
We show how different IR tasks benefit from using pre-trained word embeddings by pre-estimating representations for query and documents, or as warm start for representation learning during training.

\item[\textbf{25 mins}] Learning unsupervised representations from scratch for semantic matching --
We explain how to learn representations of words and documents in an unsupervised manner without any relevance label, that satisfy the requirements of IR problems.
\end{enumerate}


\paragraph{Learning to rank, 45 minutes (\ChristopheVanGysel, \BhaskarMitra)}

\begin{enumerate}[itemindent=25pt,leftmargin=0.15in]
\item[\textbf{10 mins}] Feature-based models for representation learning --
We explain how to train a ranker using featurized input, and how to feed the network with raw data to have it learn representations jointly with a downstream task.

\item[\textbf{15 mins}] Ranking objectives and loss functions --
We describe point-wise and pair-wise settings for the ranking task and the proper loss functions for each setting.

\item[\textbf{20 mins}] Training under different levels of supervision --
We cover how to train a neural ranker in an unsupervised way, with weak, semi- or full supervision and discuss requirements and concerns of each situation.
\end{enumerate}


\paragraph{Modeling user behavior --- 45 minutes (\AlexeyBorisov, \MaartendeRijke)}

\begin{enumerate}[itemindent=25pt,leftmargin=0.15in]
\item[\textbf{10 mins}] Biases and PGM-based click models --
    We introduce notions of bias in user behavior and
    explain how to account for them using probabilistic graphical models (PGMs).
\item[\textbf{25 mins}] Neural click models --
    We discuss weaknesses of PGM-based approaches
    and present an alternative based on recurrent neural networks.
\item[\textbf{10 mins}] Hybrid approach --
    We describe recent work
    on modeling biases in times between user actions (e.g., click dwell time)
    using ideas exploited in PGM-based and neural click models.
\end{enumerate}


\paragraph{Generating responses --- 45 minutes (\MostafaDehghani, \TomKenter)}

\begin{enumerate}[itemindent=25pt,leftmargin=0.15in]
\item[\textbf{15 mins}] Machine reading/question answering --
                 How is the se\-quence-to-sequence paradigm applied in neural QA systems.
\item[\textbf{15 mins}] Conversational IR/dialogue systems --
                 Unlike QA systems, conversational systems should maintain a state of a session.
\item[\textbf{15 mins}] General chatbots --
                 Chatbots bring their own set of challenges. How to stay consistent throughout the course of a conversation? How to maintain a persona? 
\end{enumerate}

\paragraph{Outlook --- 30 minutes (all)}

\begin{enumerate}[itemindent=25pt,leftmargin=0.15in]
\item[\textbf{15 mins}] Recent advances
\item[\textbf{15 mins}] Open research questions, current challenges
\end{enumerate}


\paragraph{Wrap up --- 15 minutes (\TomKenter)}

\begin{enumerate}[itemindent=25pt,leftmargin=0.15in]
\item[\textbf{15 mins}] Overview of material presented and conclusion
\end{enumerate}


\section{Type of support materials to be supplied to attendees}

\begin{description}
\item[Slides] Slides will be made publicly available on \url{http://nn4ir.com}.

\item[Bibliography] An annotated compilation of references will list all work discussed in the tutorial and should provide a good basis for further study.

\item[Code] Apart from the various open source neural toolkits (Tensorflow, Theano, Torch) many of the methods presented come with implementations released under an open source license. These will be discussed as part of the presentation of the models and algorithms. We provide a list pointers to available code bases.
\end{description}

\setlength{\bibsep}{0pt}
\bibliographystyle{abbrvnatnourl}
\bibliography{sigir-tutorial-nn4ir} 

\end{document}